\documentclass[10pt]{iopart}
\usepackage{amssymb,amsfonts}
\usepackage[dvips]{graphicx}

\newcommand{\msun}{{\rm\,M_\odot }}
\newcommand{\kpc}{{\rm kpc}}
\newcommand{\pc}{{\rm pc}}
\newcommand{\mm}{{\rm mm}}
\newcommand{\cm}{{\rm cm}}
\newcommand{\Jy}{{\rm Jy}}
\newcommand{\yr}{{\rm yr}}
\newcommand{\sag}{{$\mathrm{Sagittarius\,A^{*}}\,$}}
\newcommand{\sgr}{{$\mathrm{Sgr\,A^{*}}$~}}
\newcommand{\sgrs}{$\mathrm{Sgr\,A^{*}}\!\!$~}
\newcommand{\mdotfour}{${\dot{M}}_{4 \mathrm{Jy}}$\,}
\newcommand{\dF}{{{{}^{*}}\!F}}
\newcommand{\bP}{{\bf P}}
\newcommand{\bF}{{\bf F}}
\newcommand{\bU}{{\bf U}}

\newcommand{\sB}{{\mathcal{B}}}
\newcommand{\sI}{{\mathcal{I}}}
\newcommand{\sA}{{\mathcal{A}}}
\newcommand{\sJ}{{\mathcal{J}}}
\newcommand{\sN}{{\mathcal{N}}}
\newcommand{\sV}{{\mathcal{V}}}

\newcommand{\del}{{\partial}}

\newcommand{\beq}[1]{\begin{equation} #1 \end{equation}}

\newcommand{\deriv}[2]{\frac{ d #1 }{ d #2 }}
\newcommand{\pderiv}[2]{\frac{ \partial #1 }{ \partial #2 }}

\newcommand{\rg}{{G M c^{-2}}}

\newcommand{\araa}{{Ann.~Rev.~Astron.~Astrophys.}}             
\newcommand{\apj}{{Astrophys.~J.}}                 
\newcommand{\apjl}{{Astrophys.~J.~Lett.}}                
\newcommand{\apjs}{{Astrophys.~J.~Supp.}}               
\newcommand{\aap}{{Astron.\&Astrophys.}}                
\newcommand{\mnras}{{Mon.~Not.~Roy.~Astron.~Soc.}}             
\newcommand{\nat}{{Nature}}              

\newcommand{\UIUCphysics}{{Department of Physics, University of Illinois at 
Urbana-Champaign, Loomis Laboratory, 1110 West Green Street, 
Urbana, IL 61801 USA}}
\newcommand{\UIUCastro}{{Department of Astronomy, University of Illinois at  
Urbana-Champaign, 1002 West Green Street, Urbana, IL 61801 USA}}

\newcommand{\JHU}{{Department of Physics and Astronomy, 366 Bloomberg Center, 
Johns Hopkins University, 3400 North Charles Street, Baltimore, MD  21218}}
\newcommand{\IAS}{{Institute for Advanced Study Einstein Drive, Princeton, NJ 08540}}

\begin{document}

\title[Simulating the Emission and Outflows from Black Hole Accretion Disks]{Simulating 
the Emission and Outflows from Accretion Disks}

\author{
Scott C Noble$^{1,2}$,
Po Kin Leung$^{3}$,
Charles F Gammie$^{2,3,4}$,
Laura G Book$^{2}$
}

\address{1 \JHU}
\address{2 \UIUCphysics}
\address{3 \UIUCastro} 
\address{4 \IAS}

\ead{scn@jhu.edu}

\begin{abstract}

The radio source \sag (\sgr\hspace{-0.2cm}) is believed to be a hot,
inhomogeneous, magnetized plasma flowing near the event horizon of the
$3.6 \times 10^6 \msun$ black hole at the galactic center.  At a
distance of $8 \kpc$ ($\simeq 2.5 \times 10^{22} \cm$) the black hole 
would be among the largest black
holes as judged by angular size.  Recent observations  are consistent
with the idea  that the millimeter and sub-millimeter photons are
dominated by optically thin, thermal synchrotron emission.  Anticipating
future Very Long Baseline Interferometry (VLBI) observations of \sgr at
these wavelengths, we present here the first dynamically self-consistent
models of millimeter and sub-millimeter emission from \sgr based on
general relativistic numerical simulations of the accretion flow. 
Angle-dependent spectra are calculated assuming a thermal distribution of
electrons at the baryonic temperature dictated by the simulation and the
accretion rate, which acts as a free parameter in our model.  
The effects of varying model parameters (black hole
spin and inclination of the spin to the line of sight) and source 
variability on the spectrum are shown.
We find that the accretion rate value needed to
match our calculated millimeter flux to the observed flux is consistent
with constraints on the accretion rate inferred from detections of the 
rotation measure. 
We also describe the relativistic jet that is launched along the 
black hole spin axis by the accretion disk and evolves to scales 
of $\sim 10^3 \rg$, where $M$ is the mass of the black hole.

\end{abstract}

\pacs{
95.30.Jx,
95.30.Qd,
98.35.Jk,
95.85.Fm,
98.62.Mw,
98.38.Fs,
98.58.Fd,
97.60.Lf,
95.30.Sf,
02.60.Cb}

\submitto{\CQG}
\maketitle


\section{Introduction}

One of the most attractive areas of astrophysics lies at the
intersection of astronomy and gravitational physics, in the rapidly
growing observational and theoretical study of black holes in their
natural setting.  Candidate black holes are found in binary systems,
with mass $M \sim 10 \msun$, and in the nuclei of galaxies, with $M \sim
10^6$ to $10^{10} \msun$.  Intermediate-mass candidates exist but are
much less secure.  Observational facilities that operate across the
electromagnetic spectrum are gathering a wealth of new data about black
hole candidates, primarily by observing radiation from a hot, luminous
plasma deep in the object's gravitational potential 
(\cite{2006ARA&A..44...49R}, \cite{2005SSRv..116..523F}).  
In some cases this plasma is streaming outward and will be observed as 
a collimated {\it jet} at
large radius; in other cases the plasma is believed to be moving inward,
forming an {\it accretion disk}.\footnote{We use the term disk to mean
any accretion flow with angular momentum.  In some cases a
ring-like structure is directly observed (NGC 4258).}  Both the origin
of the jet and the structure of the disk are poorly understood, and new
developments in the theory of both are the subject of this paper.

The massive dark object in the center of our galaxy, which coincides
with the radio source \sgrs, is one of the most interesting black
hole candidates; from here on we will dispense with the word 
{\it candidate} for \sgr as the evidence for a black hole there 
is so strong as to make alternative models highly contrived; 
see, e.g. \cite{2006ApJ...638L..21B}.  At a distance of $R \simeq 8 \kpc$
\cite{2005ApJ...620..744G,2005ApJ...628..246E,2006ApJ...648..405B}, 
this $M \simeq 4 \times 10^6 \msun$ black hole has
a larger angular size than all candidate black holes, and therefore offers
the best opportunity for directly imaging the silhouette 
(or ``shadow'' \cite{2000ApJ...528L..13F}) of an event horizon.  However,
its remarkably small bolometric luminosity of $L\approx10^{3} L_\odot
\approx 10^{-8} L_\mathrm{edd}$---where $L_\mathrm{edd}$ is the
Eddington luminosity---provides a challenge for theoretical models.  If
one assumes that the accretion rate $\dot{M}_\mathrm{X-rays}
\simeq 4\times10^{-5} \msun \yr^{-1}$ at $r \approx 0.1\pc
\approx 5.6\times10^5 \rg$, based on a Bondi model and 
X-ray observations
\cite{2001Natur.413...45B,2003ApJ...591..891B}, holds for all $r$ and
that accretion flow is a thin disk near the black hole, then
the observed luminosity is approximately 
\beq{
L \approx 10^{-5} \left(\frac{0.1}{\eta}\right) L_\mathrm{thin}
= 10^{-5} \left(\frac{0.1}{\eta}\right) c^2 \dot{M}_\mathrm{X-rays}
\quad , 
\label{thin-disk-luminosity}
}
($\eta$ is the radiative efficiency) so either $\dot{M}$ varies with
$r$, the thin disk model is irrelevant, or both.  The spectral energy
distribution (SED) shows no sign of the multitemperature black body
distribution  expected from a thin disk (e.g.
\cite{2002luml.conf..405N}).  Recent millimeter and sub-millimeter
polarimetry observations, folded through a model of the accretion flow,
require $\dot{M} \lesssim 10^{-7}-10^{-9} \msun \yr^{-1}$
\cite{2006ApJ...646L.111M,2006astro.ph..7432M} near the hole.  All this
suggests that $\dot{M}$ drastically diminishes as $r\rightarrow0$. 

Current popular theories of \sgr fall into two categories:  jet models
and radiatively inefficient accretion flow (RIAF) models.   The former
suppose that the most luminous part of \sgr is a pair of relativistic
jets of plasma propagating perpendicular to the accretion flow that emit
via synchrotron and/or synchrotron self-Compton processes
\cite{1996ApJ...464L..67F,2000A&A...362..113F}.  The RIAF theories
suggest that the disk is quasi-spherical but rotating, and emits via
synchrotron, bremsstrahlung and Compton processes
\cite{2003ApJ...598..301Y}.  In order to account for the low luminosity,
the RIAF disk is taken to be an inefficient emitter that retains much of
its heat and maintains a geometrically thick profile.  Each of these
models are freely specified by a number of unknown parameters such as
the radius of the jet's sonic point or the fraction of heat shared
between electrons and protons in the RIAF disk.  With this considerable
freedom, each model can predict the spectrum quite well.  

These two theories neglect GR effects and do not account for dynamical
variations of the spectrum self-consistently.  General relativistic
calculations of the emission have been performed, though they have used
RIAF solutions  \cite{2006ApJ...636L.109B} or simple orbiting spheres of
hot plasma \cite{2005MNRAS.363..353B,2006MNRAS.367..905B} as sources.
They also use an isotropic (angle-averaged) synchrotron emissivity.  A
radiative transfer calculation based on Newtonian magnetohydrodynamic (MHD) 
simulation data
has been performed using the Paczynski-Witta potential to approximate
the black hole's effect \cite{2005ApJ...621..785G}, but this cannot
fully account for light-bending, gravitational redshift, and Doppler
effects, particularly if the black hole is rapidly rotating.

Here we will present the first  self-consistent optically thin
calculations of \sgrs's image and spectrum at about $\lambda=1\mm$, near
the peak of its SED.  This band of radiation is particularly interesting
since it originates near the horizon and will, consequently, be strongly
affected by the hole's curvature.  Improvements in millimeter and
sub-millimeter Very Long Baseline Interferometry (VLBI) will soon permit
features at the scale of the horizon to be resolved
\cite{2004GCNew..18....6D}; this makes the construction of accurate,
detailed models that incorporate relativistic effects even more
pressing.

Another active subject relevant to accretion disks is the study of
relativistic jets.  Whether they are black hole of a few solar masses
\cite{1999ARA&A..37..409M,2004ARA&A..42..317F} or are extragalactic and
supermassive \cite{1998ARA&A..36..539F,2006ARA&A..44..463H}, jets are
observed emanating from them.  Following the recent surge of interest in
general relativistic magnetohydrodynamic (GRMHD) simulations, 
several groups have begun to investigate the outflows
that appear spontaneously in weakly radiative accretion disk simulations
\cite{2005astro.ph..2225D,2006MNRAS.368.1561M,2006ApJ...641..103H}.  We
contribute to this body of work by presenting recent evolutions of jets
launched from geometrically thick disks.  We describe the large-$r$
scaling of the jet and explain its dependence on numerical parameters.

The outline of the paper is the following.  We describe the theory and
methodology used for our GRMHD disk simulations in
Section~\ref{sec:grmhd}.  These simulations serve as the dynamic
radiative source for our radiative transfer calculations, which are
described in Section~\ref{sec:gener-relat-radi}.  Images and spectra of
\sgr for a variety of situations are presented in
Section~\ref{sec:sgr-emission}.   Section~\ref{sec:accretion-jets}
describes our work on jets, and Section~\ref{sec:conclusion} gives a
summary.

\section{Theoretical Foundation}

In many accreting black hole systems, the inner part of the material 
flow is well explained by the ideal MHD approximation.   We employ this 
assumption in our dynamical evolutions of black hole accretion disks 
as described in the Section~\ref{sec:grmhd}.  
Emission from these simulations is calculated  via  a ray-tracing technique
described in Section~\ref{sec:gener-relat-radi}. 

\subsection{General Relativistic Magnetohydrodynamics}
\label{sec:grmhd}

We present in this section an outline  of the equations and methodology 
used to calculate accretion disk evolutions.  More thorough descriptions can 
be found in \cite{2003ApJ...589..444G,2006ApJ...641..626N}, yet we repeat 
a few points here to provide a context for the rest of the paper. 

Throughout this paper we follow standard notation \cite{mtw}.  
We work in a coordinate basis with metric components $g_{\mu\nu}$ and 
independent variables $t,x^1,x^2,x^3$.  
The quantity $n_\mu = (-\alpha, 0, 0, 0)$ is the dual of the 4-velocity of a
``normal observer'' that moves orthogonal to constant $t$ foliations of spacetime, 
where $\alpha^2 = -1/g^{tt}$ is the square of the lapse.  
Greek indices refer to all spacetime components, while Roman indices 
represent only spatial components. 
Geometrized units are used so  $G = c = 1$ unless otherwise noted. 

The GRMHD equations of motion include the continuity equation, 
\beq{
\nabla_\mu \left( \rho_\circ u^\mu \right) = 0 \quad , \label{continuity-eq}
}
the equations of local energy conservation
\beq{
\nabla_\mu {T^{\mu}}_\nu  = 0 \quad , \label{energy-conservation-eq}
}
and Maxwell's equations
\beq{
\nabla_\nu \dF^{\mu \nu}  = 0 \quad . \label{maxwell-eq}
}
Here, $\rho_\circ$ is the rest-mass density, $u^\mu$ is the fluid's 
$4$-velocity, ${T^{\mu}}_\nu$ is the MHD stress-energy tensor, and the 
Maxwell tensor $\dF^{\mu \nu}$ is the dual of the electromagnetic field 
tensor $F^{\mu \nu}$. 
The ideal MHD approximation, 
\beq{
u_\mu F^{\mu \nu} = 0 \label{ideal-mhd-eq}
}
eliminates three of the six degrees of freedom inherent in the 
electromagnetic field. The remaining degrees of freedom can be 
represented by the three non-trivial components of the magnetic 
field  in the frame of the normal observer:
\beq{
\sB^\mu \equiv - n_\nu \dF^{\mu \nu} \quad . \label{magnetic-4-vector}
}
A convenient tensor related to $\sB^\mu$ is one proportional 
to the projection of the field into a space normal to the fluid's frame: 
\beq{
b^\mu \equiv \frac{1}{\gamma} 
\left( {\delta^\mu}_\nu + u^\mu u_\nu \right) \sB^\nu \quad. 
\label{projected-magnetic-field}
}
Using these definitions, one can easily show that the MHD stress-energy 
tensor can be expressed as 
\begin{equation}
T^{\mu\nu} = \left(\rho_\circ + u + p + b^2\right) u^\mu u^\nu 
+ \left(p + {b^2\over{2}}\right) g^{\mu\nu} - b^\mu b^\nu \quad , \label{mhd-stress}
\end{equation}
where $p$ is the fluid's pressure, $u$ is the fluid's internal energy 
density, and $b^2 \equiv b_\mu b^\mu$. 
Further, one can show that the GRMHD equations of motion can take 
the following flux conservative form
\begin{equation}
\del_t \bU(\bP) = -\del_i \bF^i(\bP) + \mathbf{S}(\bP) \quad , 
\label{conservative-eq}
\end{equation}
where $\bU$ is a vector of ``conserved'' variables, $\bF^i$ are the fluxes, 
and $\mathbf{S}$ is a vector of source terms.  Explicitly, these 
are 
\beq{
\bU = \sqrt{-g} \left[ \rho_\circ u^t , {T^t}_t 
+ \rho_\circ u^t , {T^t}_j , \sB^k/\alpha  \right]^T
\label{cons-U}
}
\beq{
\bF^i = \sqrt{-g} \left[ \rho_\circ u^i , {T^i}_t + \rho_\circ u^i , {T^i}_j , 
\left(b^i u^k - b^k u^i\right) \right]^T
\label{cons-flux}
}
\beq{
\mathbf{S} = \sqrt{-g} \left[ 0 , {T^\kappa}_\lambda {\Gamma^\lambda}_{t \kappa} 
, {T^\kappa}_\lambda {\Gamma^\lambda}_{j \kappa} , 0 \right]^T \quad , 
\label{cons-source}
}
where ${\Gamma^\lambda}_{\mu \kappa}$  is the metric's associated affine 
connection coefficients. 
Note that Maxwell's equations are rewritten as the last 
three components of (\ref{conservative-eq})---also known as the induction equations---and 
a constraint equation
\beq{
\partial_i \left( \sqrt{-g} \sB^i / \alpha \right) = 0 ,  \label{divergence-constraint}
}
which must be upheld during the evolution.  Since the equations are solved
in flux conservative form, energy is conserved to machine precision.
This means that small-scale structures in the velocity
and magnetic field are erased by numerical smoothing, but that the associated
kinetic and electromagnetic energy is captured as entropy. 

We use the HARM code \cite{2003ApJ...589..444G} to evolve axisymmetric
disks on a fixed background.  Because of axisymmetry our numerical models
will fail to capture some aspects of the disk dynamics.  For example, 
axisymmetric MHD flows cannot sustain turbulence due to the anti-dynamo 
theorem and fail to properly capture the dynamics of magnetic Rayleigh-Taylor 
instabilities. 3D models will eventually be required to include these effects. 

A central, Lax-Friedrich-like flux method  similar to that proposed by
Kurganov and Tadmor \cite{2000JCoPh.160..241K} is used.  The Flux-CT
method \cite{2000JCoPh.161..605T} is used to impose the ``no-monopoles''
constraint, and the monotonized central limiter scheme is used to
reconstruct at each cell interface.  In order to calculate $\bF^i$ we
need to invert the conserved variable definitions for the primitive
variables.  This is performed using the ``2D'' method of
\cite{2006ApJ...641..626N}.  In all of the results shown here, the
equation of state 
\begin{equation}
p = \left(\Gamma - 1\right) u  \label{gamma-law-eos}
\end{equation}
is used with $\Gamma=4/3$. 
Also, we use a grid that is uniformly spaced in a slightly modified version of
the usual spherical Kerr-Schild coordinates $t,r,\theta,\phi$, which are
regular on the horizon.  The modifications concentrate numerical
resolution toward the event horizon and toward the midplane of the disk. 

Our initial data consists of a torus in hydrodynamic equilibrium
\cite{1976ApJ...207..962F}\footnote{The Fishbone-Moncrief solution
has a single key parameter $u_\phi u^t$, which is by assumption constant.
In units where $GM = c = 1$, our solution is such that $u_\phi u^t = 4.28$.}
On top of this Fishbone-Moncrief torus we add a weak magnetic field with vector
potential $A_\phi = \mathrm{Max}\left(\rho_\circ / \rho_\mathrm{max} -
0.2, 0\right)$ where $\rho_\mathrm{max}$ is the maximum of the disk's
rest-mass density.  The magnetic field amplitude is normalized so that
the ratio of gas to magnetic pressure within the disk has a minimum of
$100$.  With the addition of the field the disk is no longer strictly in
equilibrium, but because the field is weak it is only weakly perturbed.
The initial state is unstable to the MRI \cite{1991ApJ...376..214B}, so
turbulence develops in the disk and material accretes onto the black
hole.  Since HARM is incapable of evolving a vacuum, we surround the
disk in an artificial atmosphere, or ``floor'' state, with
$\rho_{\circ,\mathrm{atm}} = 10^{-4} (r/M)^{-3/2}$ and $u_\mathrm{atm} =
10^{-6} (r/M)^{-5/2}$.  Whenever $\rho_\circ$ and $u$ fall below the
floor they are artificially set to the floor.

\subsection{General Relativistic Radiative Transfer}
\label{sec:gener-relat-radi}

We consider non-polarized, optically thin emission from a thermal
distribution of electrons at wavelengths near the sub-millimeter peak in
\sgrs's spectrum.  At these wavelengths, the disk is
expected to be optically thin and thermal synchrotron emission is
expected to dominate.   We include both synchrotron and bremsstrahlung,
confirming that the former dominates.  Even though much of the disk is 
calculated to be optically thin for frequencies of interest here, 
there are small regions where absorption is important.  For this reason, 
our calculations include absorption and the radiative transfer equation 
is solved. 

Numerical methods for calculating emission in curved spacetimes have become 
more refined and sophisticated since their introduction decades ago. 
One of the first calculations including light 
bending, lensing, gravitational redshifts and Doppler redshifts
in general relativity was done by \cite{1975ApJ...202..788C} through the use
of the so-called ``transfer function,'' which maps the specific intensity 
of a luminous source to what would be observed at infinity\footnote{An 
implementation of the algorithm described in \cite{1975ApJ...202..788C} 
was written and made publicly available by \cite{1995CoPhC..88..109S}.}.  
Polarization transfer functions were implemented
in a Monte Carlo algorithm by \cite{1980ApJ...235..224C}, who
modeled polarized X-ray emission from geometrically thick clouds
around Kerr black holes.  This method was later developed 
by \cite{1990MNRAS.242..560L} to study the
effects of self-illumination on the emission from accretion 
disks \cite{1976ApJ...208..534C}.  An algebraic expression 
for the polarization transfer function in the Schwarzschild spacetime 
was derived by \cite{1991ApJ...382..125C} to study the polarization
of line emission from thin disks. Time-dependent emission from 
accretion disk hot spots was calculated using a code by 
\cite{1992MNRAS.259..569K} that ``compressed'' and stored geodesic
curves as Chebyshev polynomials 
so that many transfer calculations could be done without 
repeating the laborious geodesic integrations.  An efficient and 
somewhat complicated semi-analytical way of integrating the geodesic 
equations in Kerr spacetimes was developed by \cite{1994ApJ...421...46R}. 
The variability from indirect photons (i.e. those that follow highly 
curved geodesics) on \sgr IR emission was estimated by \cite{1997ApJS..112..423H}.   
Modern techniques for efficiently calculating optically thin line emission 
from general sources and spacetimes have been developed by a number of 
groups (e.g. \cite{1997ApJ...475...57B,2004ApJS..153..205D,2005MNRAS.359.1217B}). 
The theory of polarized radiation propagation and transfer 
through a magnetized plasma was derived and implemented in 
\cite{2003MNRAS.342.1280B,2004MNRAS.349..994B}.  This work and another
(\cite{2004A&A...424..733F}) solves the radiative transfer equations 
with absorption in covariant form.  More recently, 
a radiative transfer code that uses data from GRMHD disk simulations has 
been used to investigate the presence of quasi-periodic oscillations in 
calculated thermal radiation \cite{2006ApJ...651.1031S}.  

In our work, radiation is modeled as discrete bundles of photons that follow null 
geodesics from the disk to a ``camera'' $8 \kpc$ from \sgrs.  The geodesic 
equation is solved in first-order form:
\beq{
\pderiv{x^\mu}{\lambda} = N^\mu  \quad , \quad 
\pderiv{N_\mu}{\lambda} = {\Gamma^\nu}_{\mu \eta} N_\nu N^\eta    \quad , \quad 
\label{geodesic-eq-first-order}
}
where ${\Gamma^\nu}_{\mu \eta}$ are the connection coefficients, and
$N^a = \left(\pderiv{}{\lambda}\right)^a$ is the tangent vector along
the geodesic that is parametrized by the affine parameter $\lambda$.  As
usual 
\cite{1997ApJ...475...57B,2004ApJS..153..205D,2004ApJ...606.1098S,2006MNRAS.367..905B}, 
the geodesics are calculated in reverse from a camera pixel back
through the simulation volume.  We assume that the camera is a static
observer in our coordinates and is centered on the black hole.  From the
far ends of the geodesic, the radiative transfer equations are
integrated forward to obtain the final specific intensity values. The
initial intensities are set to zero since the bundles either start at
the event horizon or originate from past null infinity.  

The equations of general relativistic radiative transfer naturally develop 
from a generalization of Liouville's theorem to non-inertial frames 
\cite{1966AnPhy..37..487L} (see also \cite{mtw} and \cite{Mihalas-Mihalas}), 
which states that the number of photons, $d\sN$,  per phase space volume, 
$d\sV$, is invariant along the photon trajectory in vacuum:
\beq{
\deriv{}{\lambda} \frac{d\sN}{d\sV} \ \equiv \ \deriv{f}{\lambda} \ = \ 0 
\quad ; \label{Louisville-theorem}
}
here $\lambda$ is the affine parameter of the geodesic that the bundle
follows and $f$ is the photon distribution function.  It is more common
to describe the radiation field by the specific intensity $I_\nu \propto
\nu^3 f$ at frequency $\nu$, or with the invariant intensity $\sI =
I_\nu/\nu^3$.

When ionized matter is present, photons can be scattered in and out of
the bundle, converted to material degrees of freedom (absorbed) or can
be added to the bundle via spontaneous or induced emission.  In the
optically thin limit, scattering events are rare and can be ignored.
Since the rate of absorption is proportional to the bundle's
intensity and the rate of emission is not, the frame-independent
radiative transfer equation takes the simple source/sink form:
\beq{
\deriv{\sI}{\lambda} = \sJ - \sA \sI \quad ,  \label{gr-rt-eq}
}
where $\sJ$ and $\sA$ are the Lorentz invariant emissivity and absorption 
coefficient, respectively, which are related to their frame-dependent
counterparts $j_\nu$ and $\alpha_\nu$ by
\beq{
\sJ = \frac{j_\nu}{\nu^2}   \quad , \quad   \sA = \nu \alpha_\nu   \quad . 
\label{emiss-abs-defs}
}
The dimensions of $\lambda$ can be deduced by reducing
equation~(\ref{gr-rt-eq}) to the usual inertial-frame version:
\beq{
\deriv{I_\nu}{s}  = 
 j_\nu - \alpha_\nu I_\nu  \quad , \label{inertial-rt-eq}
}
where $ds = c dt_\nu$  is the path length the photon traverses over time 
interval $dt_\nu$ as measured in a frame in which the photon's frequency 
is $\nu$.  For these to be equivalent we must define $\lambda$ so that
the tangent vector appearing in the geodesic equation is
\beq{
N^\mu = \frac{c}{2 \pi} k^\mu \quad , \label{4-vector-def}
}
and $k^\mu$ is the photon wavevector.

In practice, using $\lambda$ as an integration variable leads to loss
of precision near the horizon.  We instead use 
$d\lambda^\prime = d\lambda/n(r)$ where 
\beq{
n(r) = \frac{r}{r_h} -  1  \quad ,  \label{affine-function}
}
and $r_h$ is the radius of the event horizon. 

A single time slice of the disk evolution is used to calculate $j_\nu$
and $\alpha_\nu$.  This crude approximation will be accurate where the 
matter distribution varies
slowly compared to a light crossing time, as it is in the bulk of
the disk.  It will also be accurate for observations over timescales
longer than the light-crossing time; this applies to VLBI observations 
of \sgrs.  The primitive variables,
$\left\{\rho, u, \tilde{u}^i, \sB^i\right\}$,  from the time
slice are bilinearly-interpolated at each point along the geodesic and
stored; $\tilde{u}^i \equiv \left({\delta^i}_\mu + n^i n_\mu\right)u^\mu$ 
is the spacelike velocity perpendicular to $n^\mu$. 
The step size is a tunable fraction  of the local grid spacing
so that the simulation data is well sampled.  

The interpolated data is then used to 
integrate equation~(\ref{gr-rt-eq}) using one or several
emission models.  The Lorentz invariant emissivity $\sJ$ is calculated
from the local observer's value of $j_\nu$.  We assume a thermal
distribution function for the electrons so that $\alpha_\nu = j_\nu /
B_\nu$ where $B_\nu$ is Planck's distribution.  We use an anisotropic,
angle-dependent approximation to $j_\nu$ taken from
\cite{2000MNRAS.314..183W}.   We have confirmed that this expression
yields results to an accuracy no worse than any of our other assumptions
or approximations (\cite{leung-noble-gammie-apjs}).

\section{\sgr Emission} 
\label{sec:sgr-emission}

Observations indicate that most of \sgrs's radiation
originates as optically thin emission near the event horizon at a
wavelength of $\lesssim1\mm$ (e.g. \cite{2006astro.ph..7432M}).  Even if
the accreting plasma has very little angular momentum, it is expected to
have circularized in this region.  If the magnetic field is weak at
large radii, it is expected to be amplified to near equipartition with
the disk's internal energy via the magnetorotational instability (MRI)
\cite{1991ApJ...376..214B}.  This makes previous  accretion disk
simulations  \cite{2004ApJ...611..977M} suitable for our study since
they yield statistically steady flows within $r\lesssim12M$.  

The electrons are assumed to follow a thermal distribution which is
consistent with modern models that find a power-law distribution of
electrons is needed to simultaneously match radio and X-ray
observations, but that thermal electrons are the dominant emitters at
sub-millimeter/millimeter wavelengths
\cite{2003ApJ...598..301Y,2002A&A...383..854Y}.  Further, we assume that
the electrons and ions are at the same temperature, 
although some successful models of \sgrs's spectrum and variability
assume a two-temperature flow
\cite{2003ApJ...598..301Y,2005ApJ...621..785G}.  
Cooling times for
synchrotron and bremsstrahlung emission are long compared with the
inflow time in our model, as expected for the matter near \sgrs.  
Numerically integrated optical depths indicate
that that the disk is everywhere optically thin (i.e. optical depth is
less than unity) when $\lambda \lesssim 1 \mm$.  We can therefore
neglect the radiation's effect on the the GRMHD simulations for these
wavelengths.

\begin{table}[htb]
\caption{\label{table:disk-parameters} 
Parameters for the accretion disk evolutions 
used to model \sgr emission.  All quantities are given in 
geometrized units unless explicitly stated 
otherwise.  The radii $r_1$, $r_2$ and $r_\mathrm{ISCO}$ are---respectively---the 
radius of the inner edge of the equilibrium torus at $t=0$, radius
of the pressure maximum at $t=0$, and the radius of the ISCO for the given 
spacetime. 
\mdotfour is the accretion rate  resulting 
in a flux density of $4 \mathrm{Jy}$ at Earth; this is used to set the scale of the
rest-mass density.  
The quantities $\langle\dot{M}\rangle$, 
$\langle\dot{E}\rangle$, $\langle\dot{L}\rangle$ 
are---respectively---the average accretion rates of the rest-mass, energy, and 
angular momentum, taken over $1100M < t < 1500M$.
}
\begin{indented}
\item[]
\begin{tabular}{cccccccc}
\br
 $a_*$   
&$r_1$
&$r_2$
&$r_\mathrm{ISCO}$
&\mdotfour
&$\langle\dot{M}\rangle$
&$\langle\dot{E}\rangle/\langle\dot{M}\rangle$
&$\langle\dot{L}\rangle/\langle\dot{M}\rangle$ \\
$(M)$
&$(M)$
&$(M)$
&$(M)$
&$(10^{-9} \msun \yr^{-1})$
&&&\\ 
\mr
$0.0    $    &$6.4$     &$15.05$     &$6.00$   &$7.34$  &$0.88$   &$0.95$   &$3.01$\\
$0.5    $    &$6.0$     &$13.02$     &$4.23$   &$3.60$  &$0.75$   &$0.94$   &$2.58$\\
$0.75   $    &$6.0$     &$12.35$     &$3.16$   &$2.05$  &$0.41$   &$0.90$   &$2.15$\\
$0.88   $    &$6.0$     &$12.10$     &$2.48$   &$1.15$  &$0.30$   &$0.89$   &$1.96$\\
$0.94   $    &$6.0$     &$12.00$     &$2.04$   &$0.82$  &$0.23$   &$0.86$   &$1.62$\\
$0.97   $    &$6.0$     &$12.00$     &$1.75$   &$1.23$  &$0.33$   &$0.86$   &$1.65$
\end{tabular}
\end{indented}
\end{table}

The degrees of freedom of our model include the spin of the black hole
($a_*$), the accretion rate ($\dot{M}_\mathrm{scale}$), the inclination
(angle $i_\mathrm{inc}$ between the black hole angular momentum vector
and the line of sight to the black hole), and the time at which we make
the image ($t_\mathrm{pic}$).  The spin and inclination are unknown for
\sgrs, though a recent periodicity in X-ray flux that has
been seen may be evidence of $a_* \gtrsim 0.22$
\cite{2006astro.ph..4337B}.  
Disk simulations with different $a_*$ and
initial disk distributions are used and are tabulated in
Table~\ref{table:disk-parameters}.  Each simulation used $256\times256$
cells with the algorithm described in Section~\ref{sec:grmhd}.  
This resolution proves to be sufficient for our purposes since 
higher resolution simulation data (using $512\times512$ and $1024\times1024$ 
cells) produced differences easily accounted for by time variability. 
We set $r_1$ and $r_2$---the radii of the inner torus edge and the pressure
maximum---so that all tori have similar shapes and sizes initially.  The
density is scaled until the flux density at $1\mm$ matches an
observationally determined $4\Jy$ \cite{2006ApJ...640..308M};
this yields an accretion rate we will call \mdotfour.  Finally, we 
find insignificant dependence on the simulation's floor model 
in our emission calculations since its flux is always many orders
of magnitude smaller than that of the rest of the flow.

It is interesting to note that \mdotfour are always consistent with
the observational limits $\dot{M} \lesssim 10^{-7}-10^{-9} \msun
\yr^{-1}$ \cite{2006ApJ...646L.111M,2006astro.ph..7432M}
obtained by folding measurements of the rotation measure through a
model for Faraday rotation within the accretion source.
Because they use a RIAF-like model the agreement with the simulations
may in part be coincidental; it would be very interesting to see
self-consistent calculations of the rotation measure from the numerical
simulations.

\begin{figure}
\begin{tabular}{ll}
\begin{tabular}{ll}
\includegraphics[scale=.3]{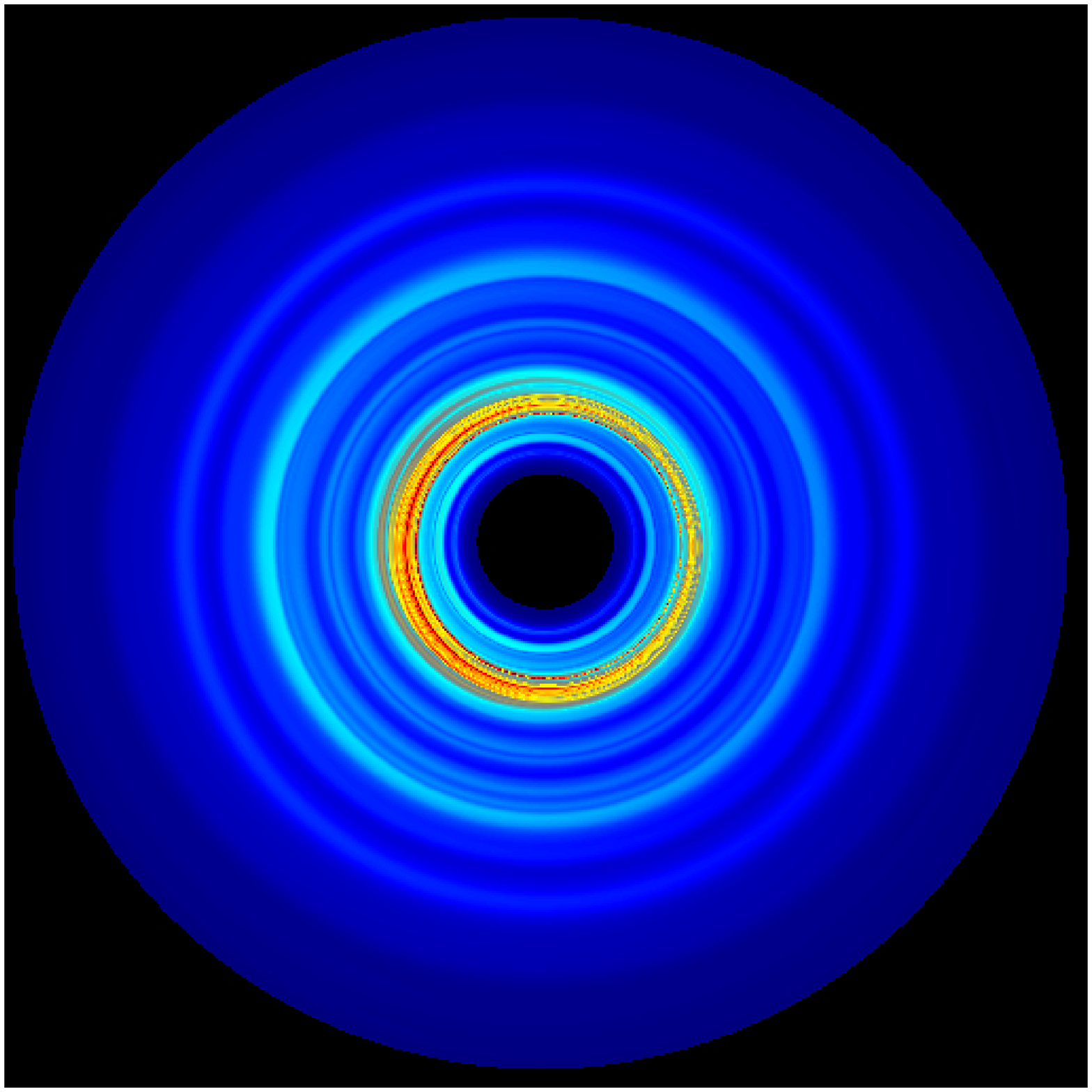}
&\includegraphics[scale=.3]{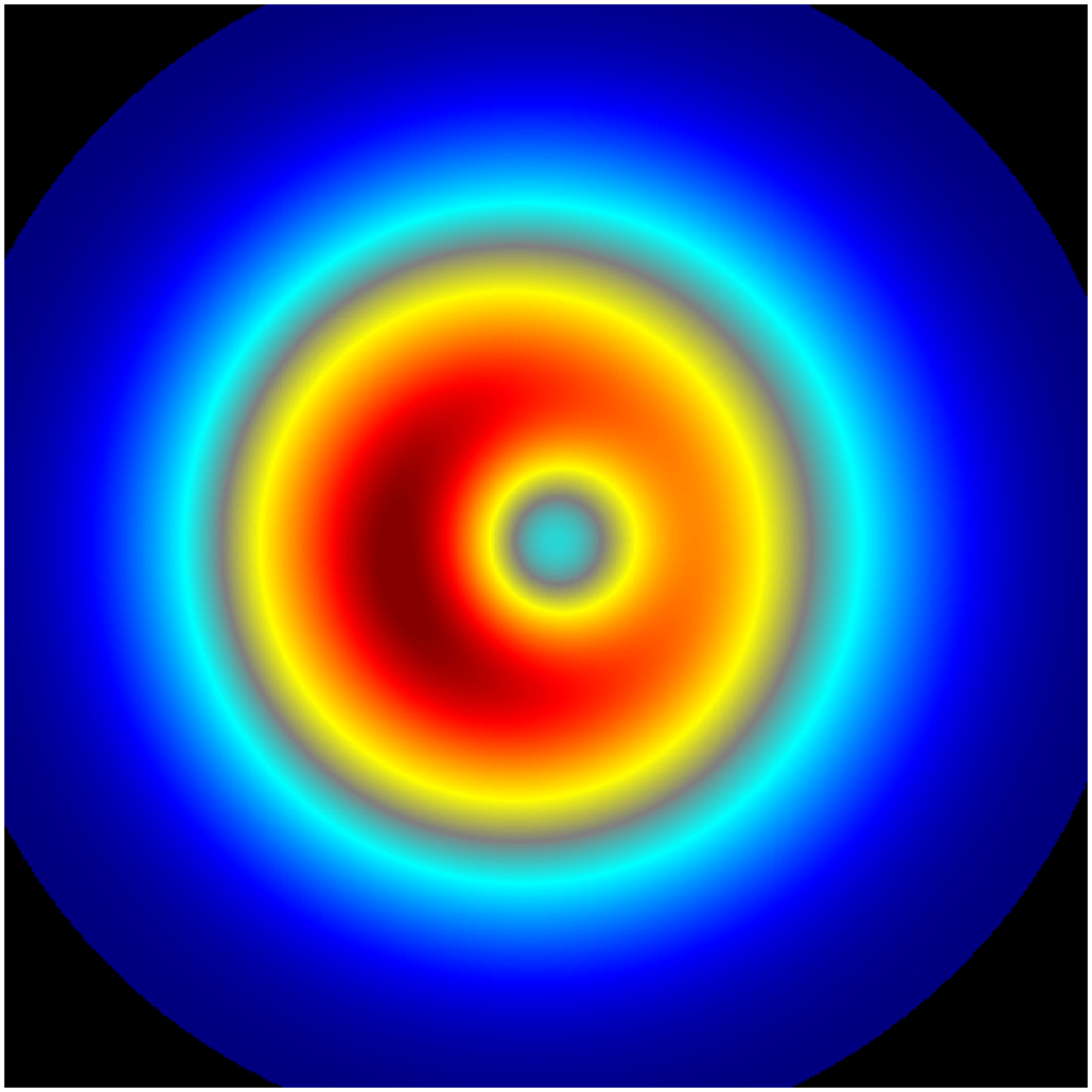}  \\
\includegraphics[scale=.3]{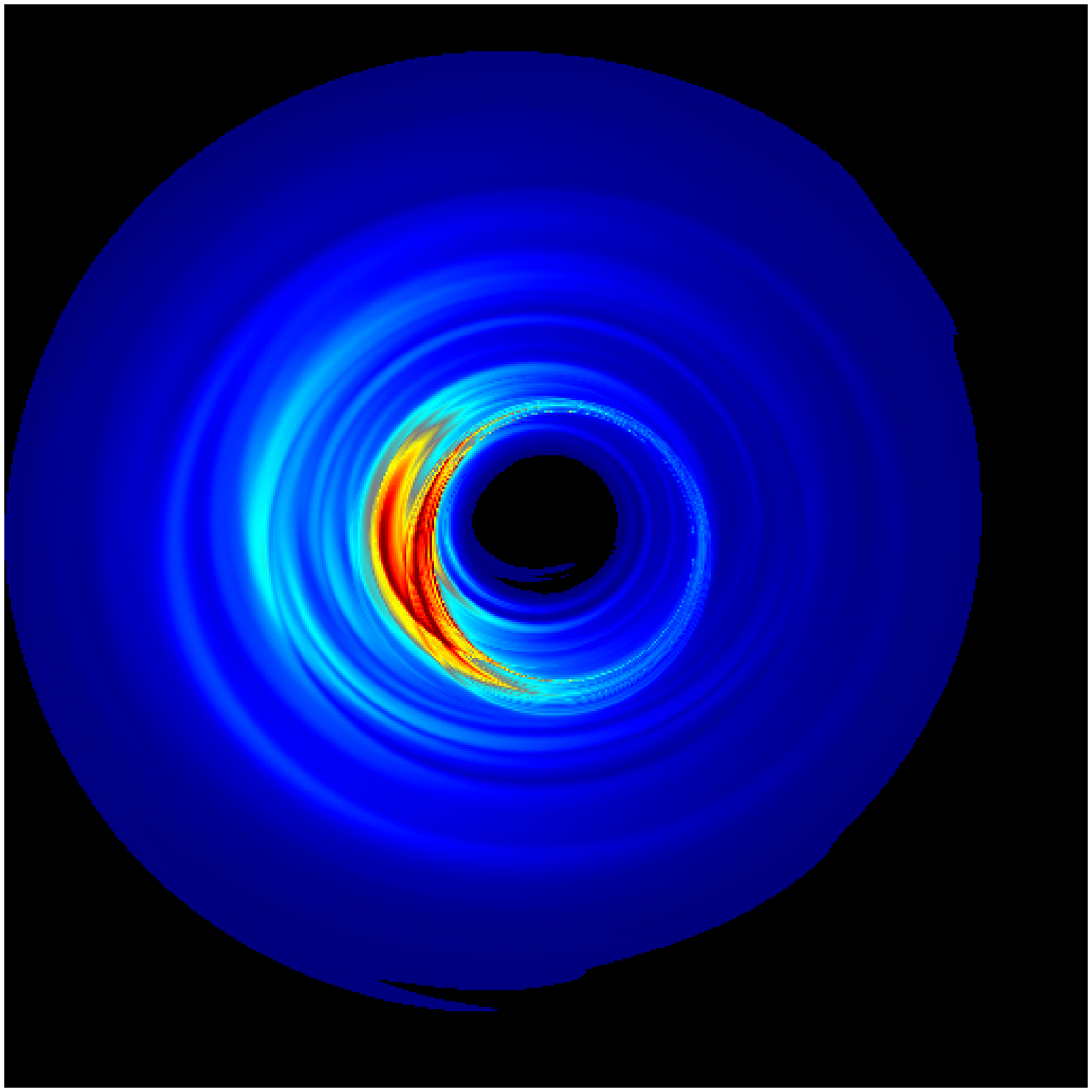}
&\includegraphics[scale=.3]{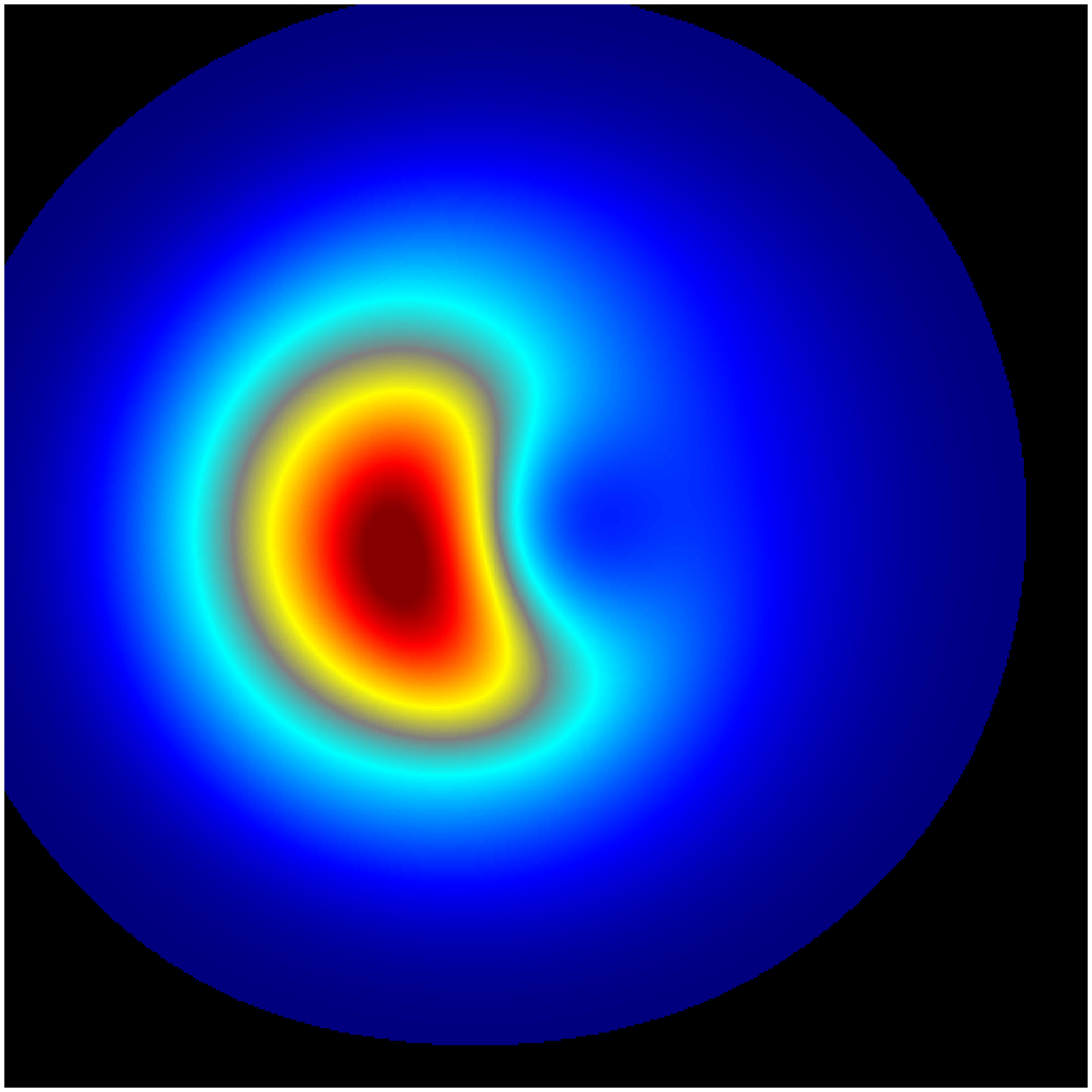}  \\
\includegraphics[scale=.3]{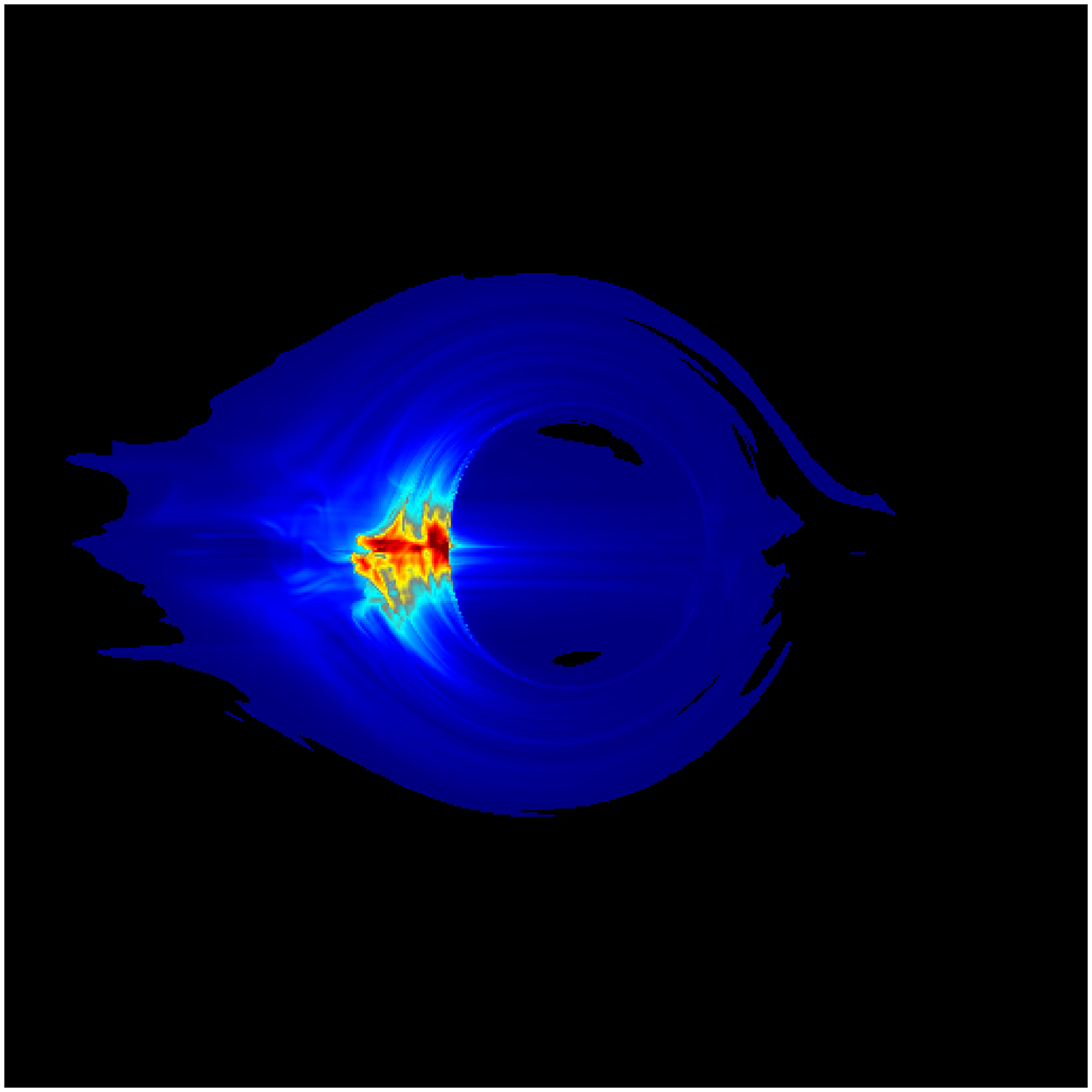}
&\includegraphics[scale=.3]{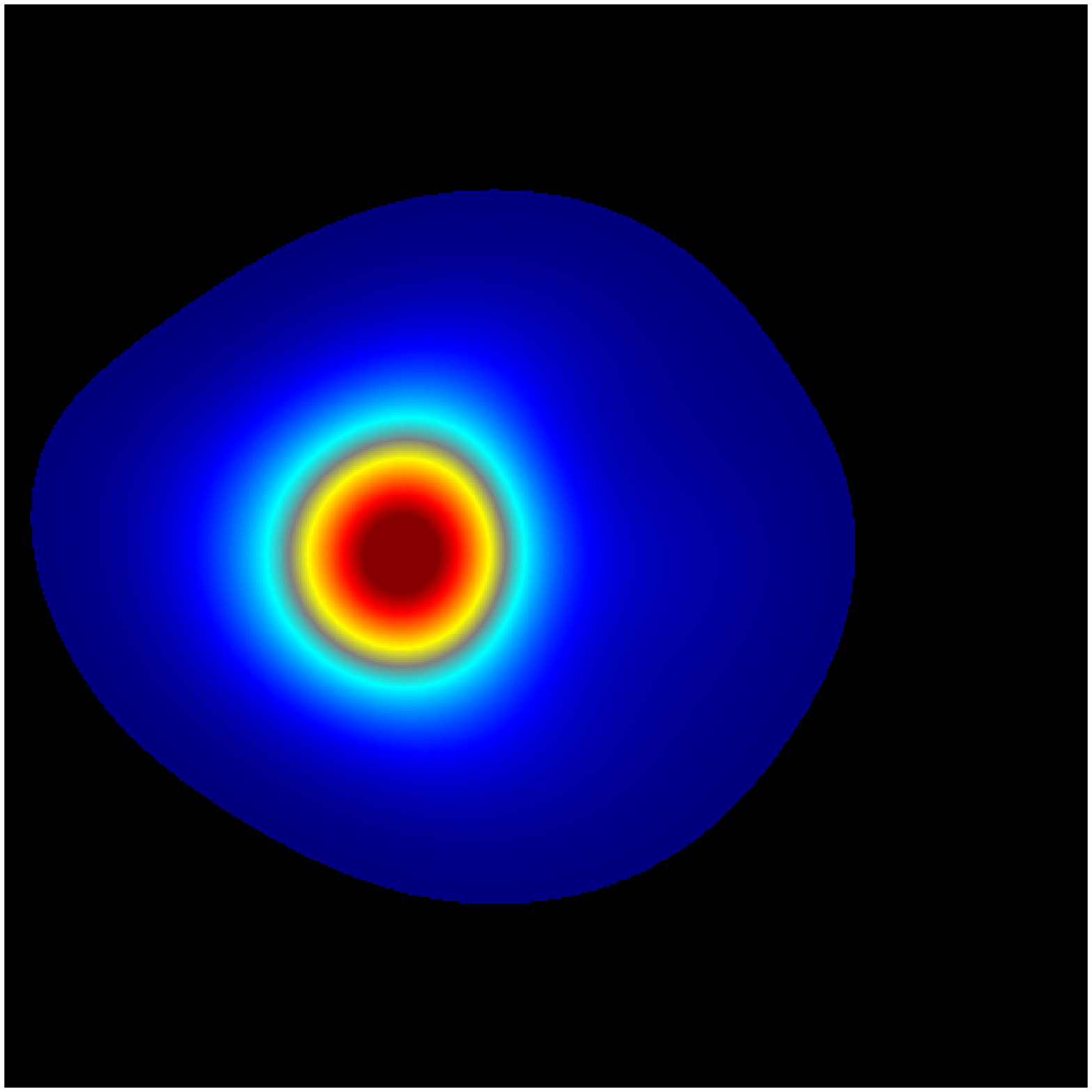} 
\end{tabular}
&\begin{tabular}{ll}
\\
\includegraphics[scale=1]{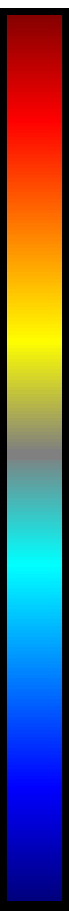}
\end{tabular}
\end{tabular}
\caption{
Images of the accretion disk viewed at a wavelength of $1\mm$ seen at
inclination angles of $5^\circ$ (top), $30^\circ$ (middle) and
$90^\circ$ (bottom).  Each frame shows a view $40M$ wide in the plane of
the singularity.  Frames in the left column are ``infinite'' resolution
images, while those in the right column have been convolved with a
symmetric Gaussian beam to simulate a $8000\mathrm{km}$ baseline VLBI
observation.  The linear colour map used is shown at the right of the images.
Each image has been scaled by its maximum intensity for illustrative purposes.
\label{fig:images}
}
\end{figure}

An outstanding concern for future millimeter/sub-\mm VLBI experiments is
whether there will be any observable effect from the black hole's
curvature.  For this purpose, we present images of a single snapshot
($t_\mathrm{pic}=1250M$) of a simulation ($a_*=0.94$) at
$\lambda=1\mm$ for different inclination angles in
Figure~\ref{fig:images}.  The raw images, in which each pixel represents
a unique ray, are shown next to their convolved counterparts.  The
convolution is performed using a circular Gaussian beam to simulate the
appearance of an image taken with VLBI  using a baseline
$8000\mathrm{km}$ at $\lambda = 1\mm$
\cite{2004GCNew..18....6D,2000ApJ...528L..13F}.  As expected, we find
that the brightest regions of the disk lie in the inner equatorial
region of the flow where $u$ and $b^2$ are largest.  The part of the
disk approaching the camera is brightest because of relativistic
beaming.  The brightest region is especially interesting since most, if
not all, of the geodesics that pass through it originate near the
horizon and orbit the black hole multiple times before reaching the
camera.  Please note that the asymmetry seen in the $i_\mathrm{inc}=5^\circ$ 
images is expected since the disk is slightly inclined to the viewer.  
Even though it is much more noticeable in the convolved image, the asymmetry is 
also present in the high resolution image.

\begin{figure}
\centerline{
\includegraphics[scale=.4]{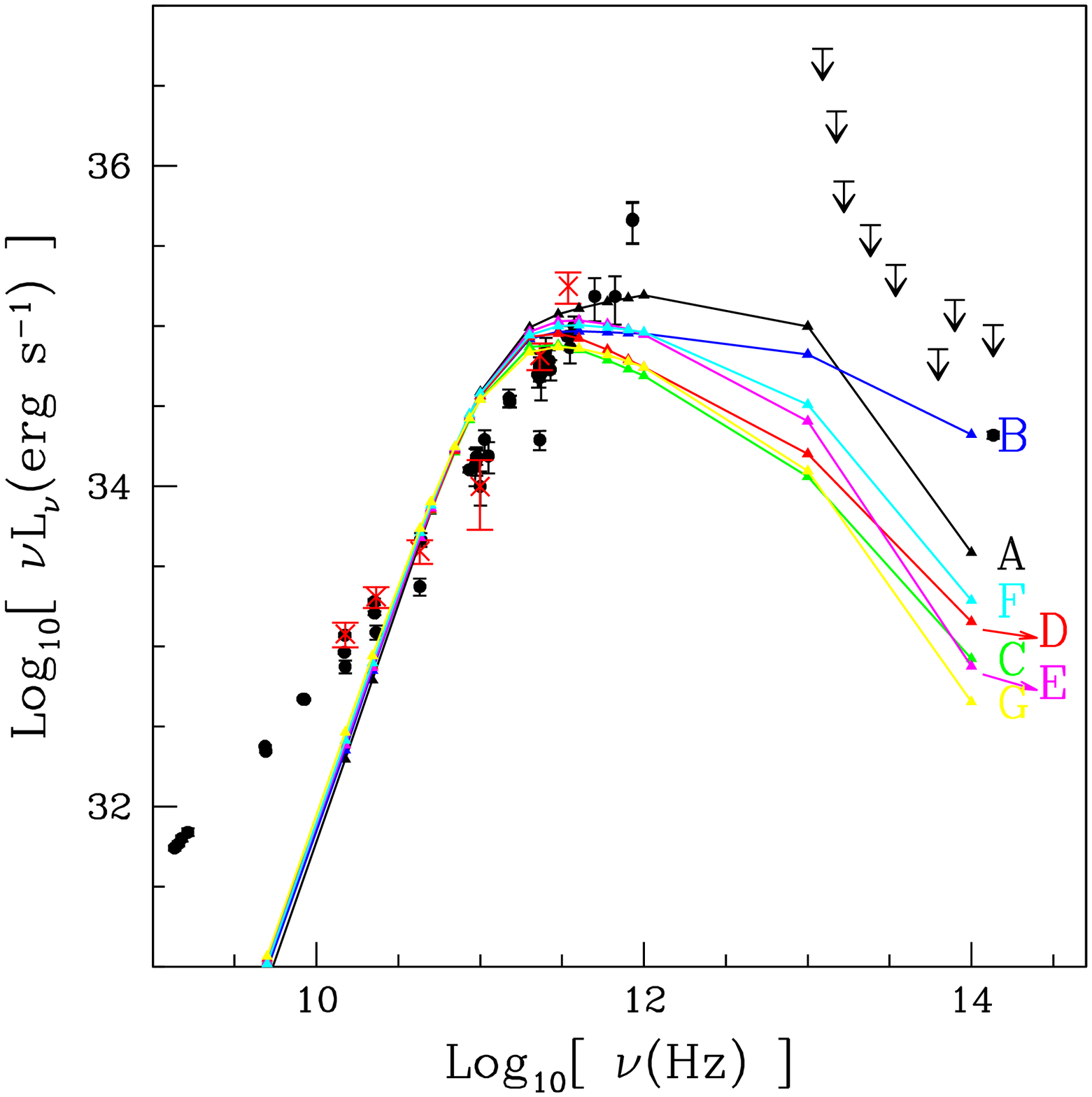}
}
\caption{Spectra taken at $i_\mathrm{inc}=30^\circ$ using snapshots of 
the $a_*=0.94$ disk at different points along its evolution.
Lines A-G respectively represent 
$t_\mathrm{pic}=1150M, 1250M, 1326M, 1434M, 1500M, 1560M, 1666M$.
 \label{fig:spectram-times}}
\end{figure}

The black hole silhouette is obvious in the raw images at all inclinations,
though may only be observable in practice if $i_\mathrm{inc} \lesssim
30^\circ$.  This does not necessarily mean that other observables---such
as variability and polarization fraction---are not sensitive to
relativistic effects at other inclinations.  

\begin{figure}
\centerline{
\includegraphics[scale=.4]{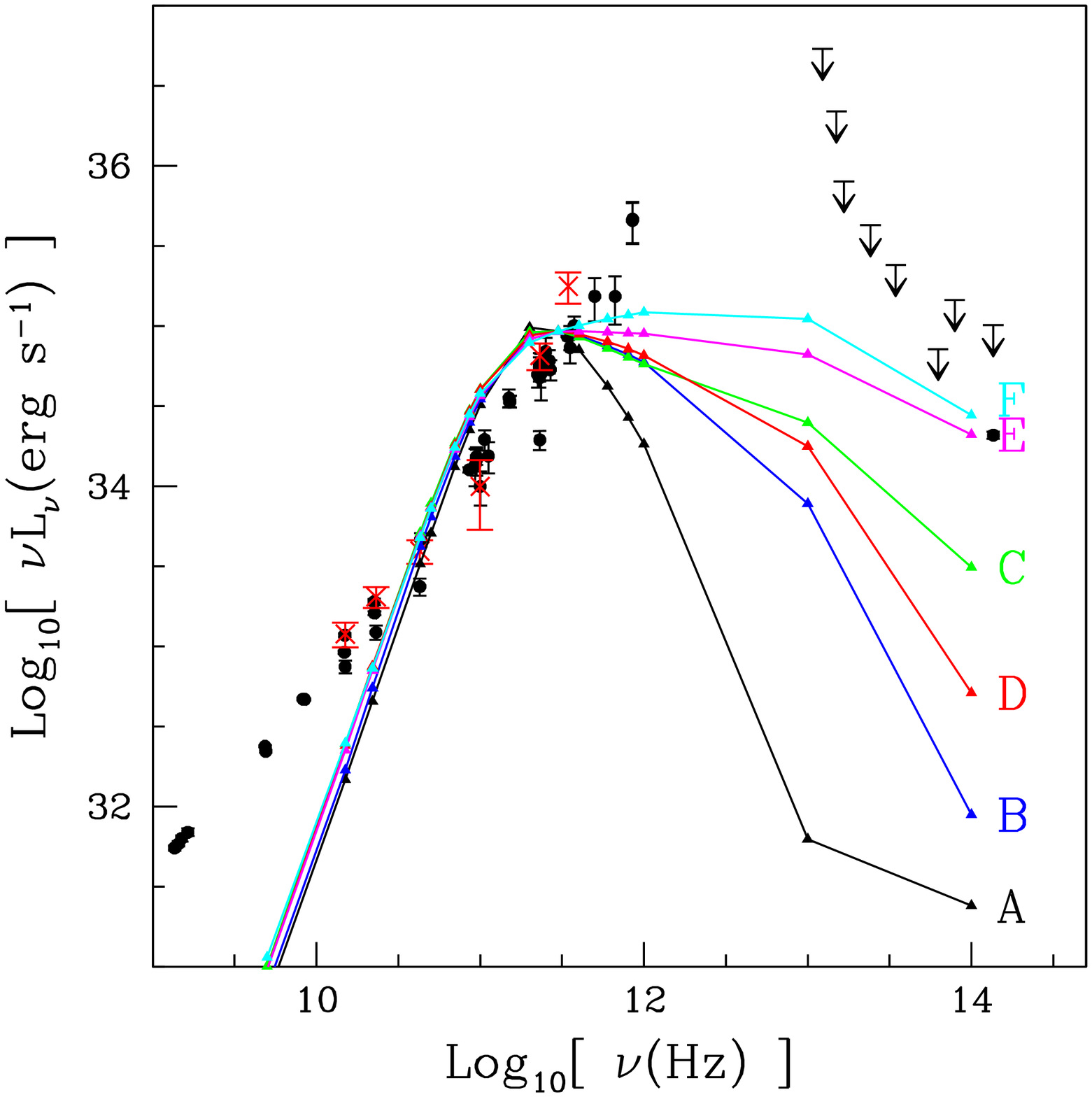}
}
\caption{Spectra taken at $i_\mathrm{inc}=30^\circ$ and $t_\mathrm{pic}=1250M$, but 
using simulation data from evolutions with different black hole spins. 
Lines A-F respectively represent $a_* = 0, 0.5, 0.75, 0.88, 0.94, 0.97$.
 \label{fig:spectram-spins}}
\end{figure}

\begin{figure}
\centerline{
\includegraphics[scale=.4]{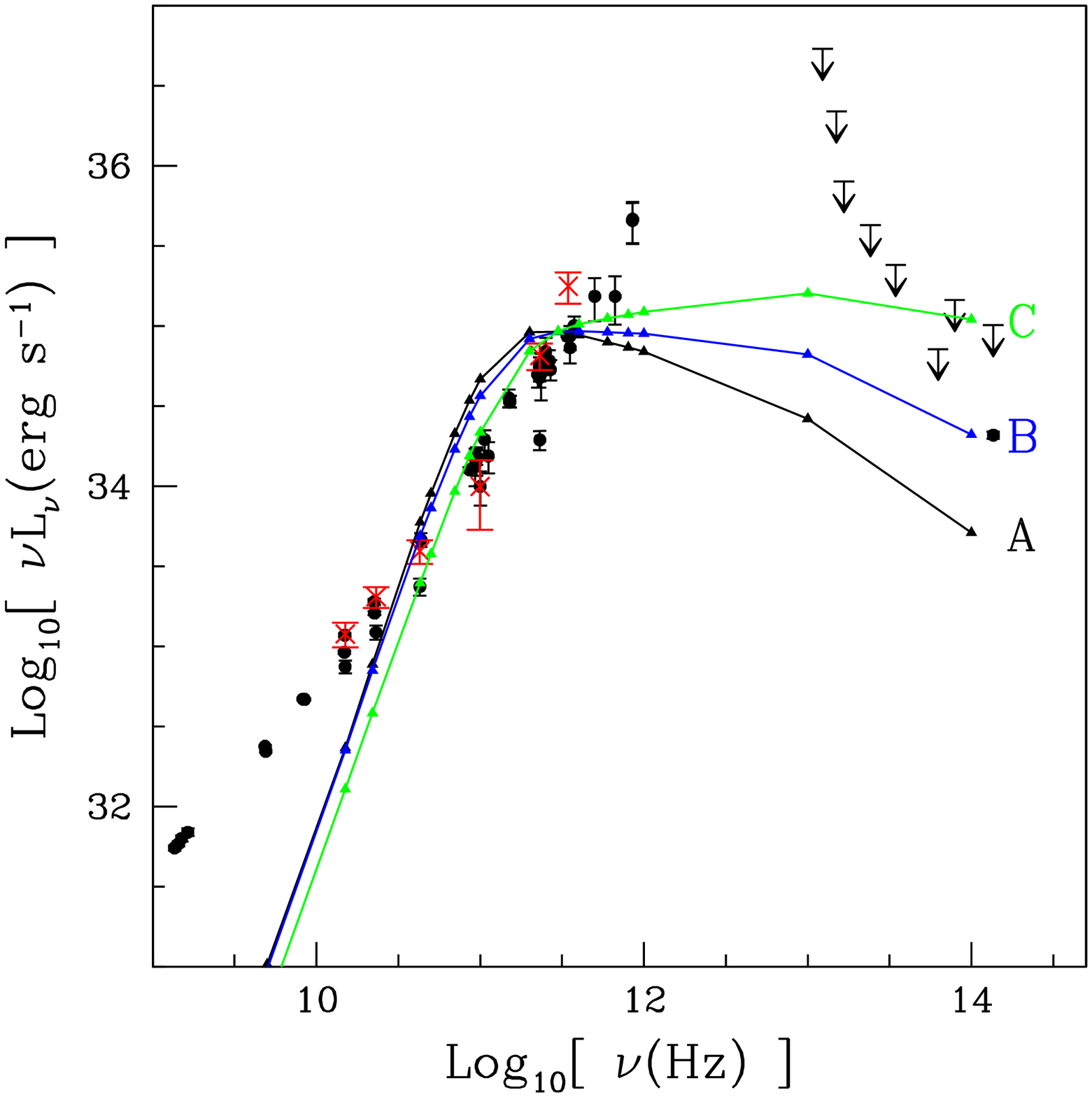}
}
\caption{Spectra taken at $t_\mathrm{pic}=1250M$ using $a_*=0.94$ simulation data
at different $i_\mathrm{inc}$. 
Lines A-C respectively represent $i_\mathrm{inc} = 5^\circ, 30^\circ, 90^\circ$. 
 \label{fig:spectram-incl}}
\end{figure}

We have also calculated spectra for a survey over $t_\mathrm{pic}$,
$a_*$, and $i_\mathrm{inc}$, shown in
Figures~\ref{fig:spectram-times}~-~\ref{fig:spectram-incl}.  A standard
model was used for comparison: $t_\mathrm{pic} = 1250M$, $a_* = 0.94$,
$i_\mathrm{inc}=30^\circ$.  The filled circles with error bars in these
plots represent observed flux values of \sgr during quiescence
\cite{1997ApJ...490L..77S,1998ApJ...499..731F,2002ApJ...577L...9H,2003ApJ...586L..29Z,2006ApJ...640..308M,2006ApJ...646L.111M}.
The red exes are found from flux measurements during flare events
\cite{2003ApJ...586L..29Z}, and the arrows indicate upper limits at
NIR/IR wavelengths \cite{1997ApJ...490L..77S}.   Error bars indicate the
measured errors quoted in the references.  We calculate $L_\nu$ assuming
isotropic emission. 

Our calculations are most relevant in the vicinity of
$\nu\simeq3\times10^{11}\mathrm{Hz}$.  At lower frequencies the emission
likely arises from plasma outside the computational domain, and so we
cannot model it.  The absence of this material may explain why the
calculated luminosities are too large at $\nu \approx
10^{11}\mathrm{Hz}$.   For $\nu \gtrsim 10^{13}\mathrm{Hz}$, both
Compton scattering and direct synchrotron emission from a power-law
distribution of electrons may be important; these effects are not
modeled here.

Temporal variations in the spectrum are small at
$\nu\simeq10^{11}\mathrm{Hz}$.  Near the peak, the variation is
comparable to current observational sensitivities and may be able to
account for some flares.  The complexity of the radiative transfer
calculation is evident in the nonuniformity of the time variability with
frequency.  

The sensitivity of the spectrum to the time slice used to calculate the
spectrum is dwarfed by the sensitivity of the spectrum to the choice of
black hole spin $a_*$.  We find a fairly uniform trend of increasing
bolometric luminosity with spin (while holding the $1\mm$ flux at $4\Jy$);
the $a_*=0.75$ and $a_*=0.88$ cases break this trend, but this may just
be the result of a temporary fluctuation.  The variation of the spectrum
with $a_*$ may be  attributable to an increase in relativistic beaming
with spin, and an increase in temperature and magnetic field strength
near the horizon with spin.  The latter is not as strong an effect as
the former since \mdotfour at $a_*=0$ is about $7$ times the value at
$a_*=0.97$.  

The SED dependence on $i_\mathrm{inc}$ is the most telling in that
NIR/IR upper limits likely rule out edge-on disks with large $a_*$.
Assuming the same trend in $i_\mathrm{inc}$ at comparable values of
$a_*$, the NIR/IR upper limits constrain our models to have
$i_\mathrm{inc}\lesssim30^\circ$ for $a_*\gtrsim0.88$ and any
inclination for smaller spins.  Coincidentally,
$i_\mathrm{inc}\lesssim30^\circ$ is also the range in inclination angle
that provides the best chance at observing the black hole's silhouette at
$\lambda=1\mm$.  Recently, the variability 
seen in flux and polarization angle at NIR wavelengths has been 
shown to be consistent with emission from an orbiting hot spot 
and ring inclined at $\gtrsim 35^\circ$\cite{2006A&A...460...15M}.  
Our results with their constraint on inclination angle then suggest
that $a_*\lesssim0.88$.  Future fits from numerical models will 
more strongly constrain the inclination and spin.

\section{Accretion Disk Jets} 
\label{sec:accretion-jets}

Jets are almost always seen in our accretion disk simulations.  They 
produce very little emission in the \sgr models considered earlier
because they are nearly empty of mass at small radii.  As they 
mix with the surrounding material present at larger radii, the jets may 
become more luminous.  Since they play no significant role in our 
calculation of \sgrs's emission, they are considered separately in this 
section. 

Our simulations of jets launched from accretion flows 
extend to distances of $r \sim 10^3 M$.  They are not dependent 
on the inner radial boundary condition since it is causally disconnected
from the rest of the numerical domain (i.e. it lies within the event horizon), 
and we terminate their evolution before any matter reaches the outer boundary. 
These outflows are generated spontaneously by a combination of forces very close 
to the black hole.
They are interesting because they are easily observable, and their large
observed Lorentz factors were used as one of the first arguments for
relativistically strong gravitational fields in the engine that drives
them.

Early attempts at studying jets with HARM were plagued by instabilities
that we have since cured by using a new set of coordinates that 
improve the resolution along the symmetry axis.  Specifically, we use
\beq{
r = e^{x^1} \quad , \quad 
\theta = \pi x^2 + \frac{2 h_2 }{\pi} \sin(2 \pi x^2)  
\arctan\left[s\left(x^1_0 - x^1\right)\right] , 
\label{new-mks}
}
which are different from the modified Kerr-Schild coordinates described in 
\cite{2003ApJ...589..444G}. 
Here $x^1_0$ is a transition radius where the grid begins to focus
toward the axis, the parameter $s$ controls how quickly this transition
is made, and $h_2$ controls the strength of the focusing.  We set
$x^1_0=\log(40)$, $s=2$, and $h_2=0.35$ so that the cells become more
focused along the axis at a radius beyond the bulk of the disk.

\begin{figure}
\centerline{
\includegraphics[scale=.4]{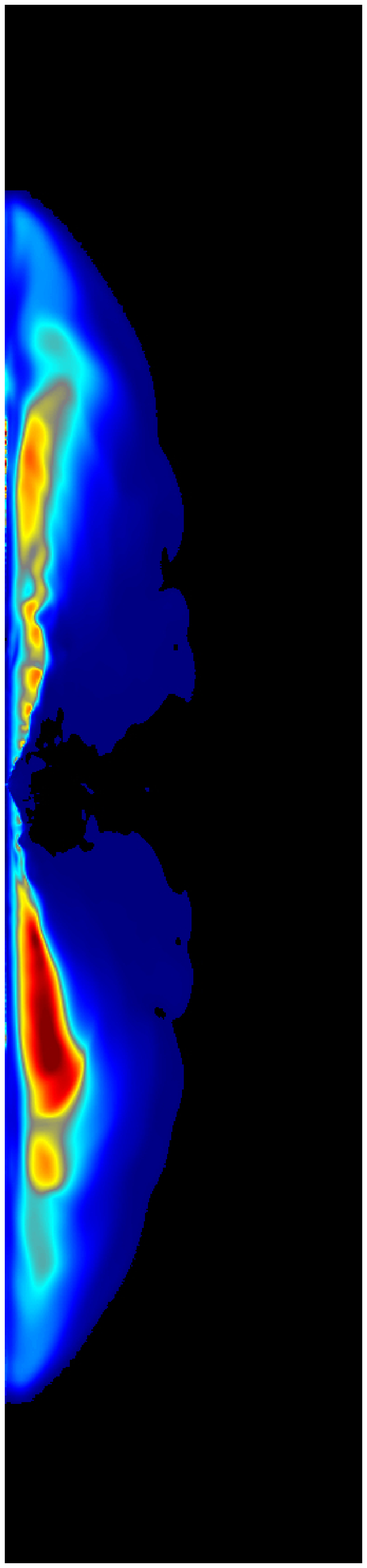}
\includegraphics[scale=.4]{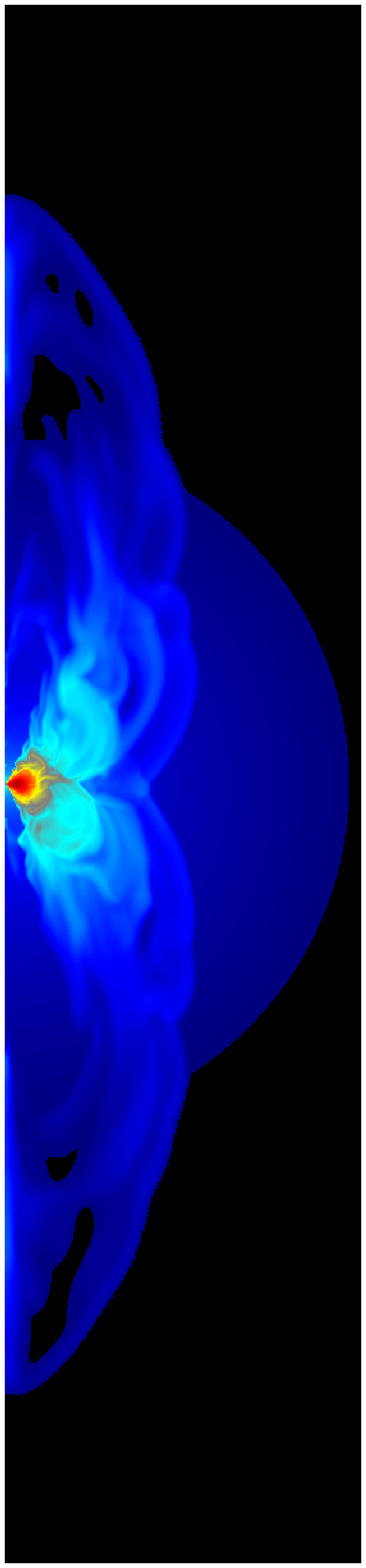}
\includegraphics[scale=.4]{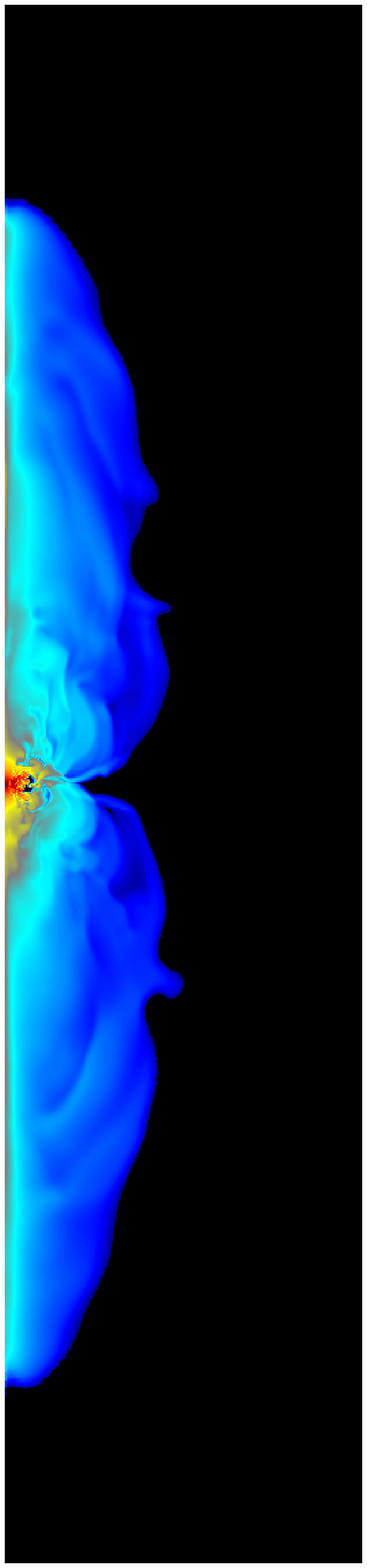}
\includegraphics[scale=.4]{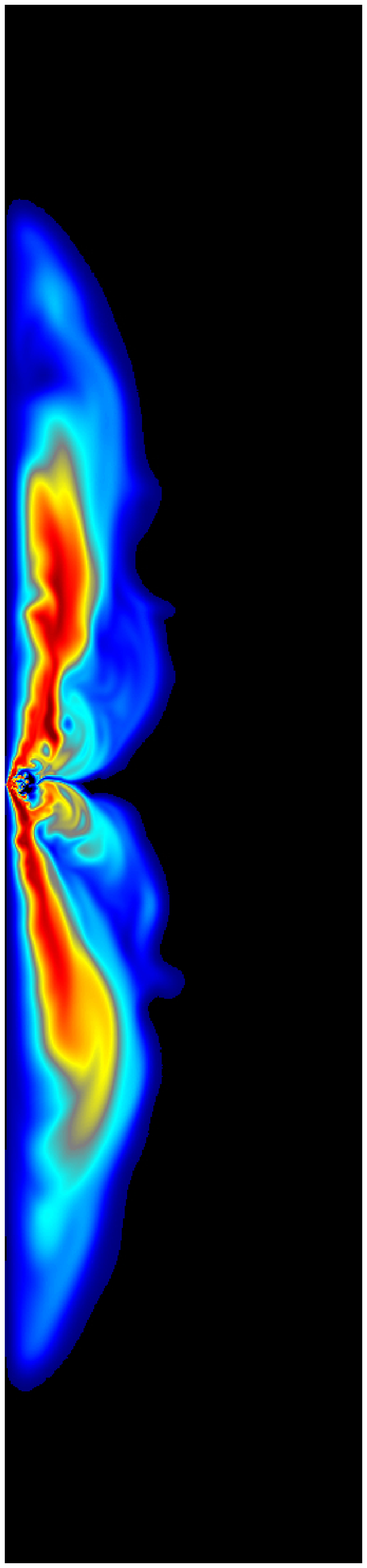}
}
\caption{From left to right are snapshots of 
$\gamma$, $\rho_\circ$, $b^2$ and $A_\phi$---whose isosurfaces follow poloidal magnetic
field lines---at $t_\mathrm{pic}=1500M$ for a run using $568\times256$ cells. The height 
of each image is $2000M$.  White (red in the colour version) represents the 
maximum of the colour scale, and black (blue in the colour version) the minimum. 
Logarithmic colour scales are used for $\rho_\circ \in \left[10^{-8},1\right]$ 
and $b^2 \in \left[10^{-10},10^{-3}\right]$.  Linear colour scales are used for
$\gamma \in \left[1,3.5\right]$ and $A_\phi \in \left[0,0.08\right]$. 
 \label{fig:jet-images}}
\end{figure}

\begin{figure}
\centerline{
\includegraphics[scale=.4]{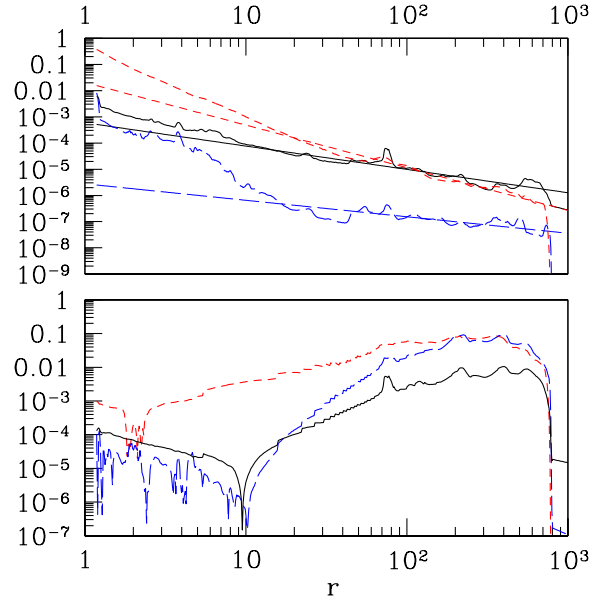}
}
\caption{
Profiles of $\rho_\circ$ (black solid line), $p$ (blue long dashes)
and $b^2/2$ (red short dashes) are shown in the top figure. 
The radial dependence of $\dot{m}$  (black solid line),  $\eta_M$ (blue long dashes), 
and $\eta_{EM}$ (red short dashes) are shown in the bottom plot.  All quantities 
are calculated at $t=1500M$ using an opening angle of $15^\circ$ from the 
axes.  \label{fig:jet-profiles}}
\end{figure}

Using these coordinates we performed a run using $568\times256$ cells
(more cells are needed to extend the grid radially).  The same initial
conditions were used as in the $a_*=0.94$ run of
Table~\ref{table:disk-parameters}.  We present here data from $t=1500M$,
which is near the end of the period of time-steady accretion.  At this
point the jet has reached $r\simeq800M$.  Figure~\ref{fig:jet-images}
shows snapshots of the Lorentz factor ($\gamma \equiv \alpha u^t$),
rest-mass density ($\rho_\circ$), magnetic field density ($b^2$) and the
toroidal component of the electromagnetic potential ($A_\phi$).  Notice
that poloidal magnetic field lines follow isosurfaces of $A_\phi$.  The
jet is magnetically-dominated and remains well collimated for at least
the first $10^3 M$.   The jet seems to be driven by Poynting flux near
the poles, and---further away from the axis---by a relativistic wind
driven both thermally and centrifugally.

The jet is relativistic, occasionally reaching $\gamma\sim 10$.  But the
maximum $\gamma$ reached is sensitive to the magnitude and profile of
the floor.  To quantify this dependence, we performed three runs with
$256\times128$ cells using the floor profiles: $\rho_{\circ
\mathrm{min}} \in \left\{ 0.2 , 1 , 5 \right\} 10^{-4}\, r^{-3/2}$,
$u_{\mathrm{min}} \in \left\{ 0.2 , 1 , 5 \right\} 10^{-6}\, r^{-5/2}$.
The maximum values of $\gamma$ averaged over $0 < \theta < 15^\circ$ at
$t=1500M$ are $6$, $2.5$, and $2$---respectively---for these floors.
Differences in the density and pressure profiles are also present since
the floor is reached throughout the polar regions in all these
instances.  The lowest floor in this set is close to the stability limit
for HARM.  By using a higher-order reconstruction method,
\cite{2006MNRAS.368.1561M} found that HARM can be extended to reliably
evolve similar outflows with a floor which is sufficiently steep and that 
the floor is reached only near the base of the jet.  Comparisons between
these results and ours further indicate how the floor affects the jet;
for example, our $\gamma$  is almost a factor of $2$ smaller. 

The radial profiles of $\rho_\circ$, $p$ and $b^2/2$ averaged over
$\Delta\theta=15^\circ$ from both poles are shown in
Figure~\ref{fig:jet-profiles}.  The thinner lines in the figure are
power-law fits to the data at $r>10M$.  We find 
\beq{
\rho_\circ \sim r^{-0.9} \quad , \quad 
p \sim r^{-0.6} \quad , \quad
b^2 \sim r^{-1.6} \quad . \label{power-law-fits}
}
The $\rho_\circ$ fit agrees with that seen by \cite{2006MNRAS.368.1561M}
for $r<120M$, but is much shallower than what they see for $r>120M$.
This is most likely attributable to the jet accumulating matter from our
larger floor. 

Also plotted are the jet luminosity and mass flux efficiencies.  The
matter and electromagnetic components of the jet's luminosity efficiency
are 
\beq{
\eta_M = \frac{2 \pi}{\epsilon \langle\dot{M}\rangle} 
\int_{d\theta_\mathrm{jet}} 
\left( -{{\hat{T}}^r}{}_t - \rho_\circ u^r \right) \sqrt{-g}  d\theta 
\quad , \label{matter-lum-efficiency}
}
\beq{
\eta_{EM} = \frac{2 \pi}{\epsilon\langle\dot{M}\rangle} 
\int_{d\theta_\mathrm{jet}} -{{\tilde{T}}^r}{}_t \, \sqrt{-g}  d\theta 
\quad , \label{em-lum-efficiency}
}
where $d\theta_\mathrm{jet}$ represents the first $15^\circ$ from both 
poles, ${{\hat{T}}^r}{}_t$ is the matter component of ${T^r}_t$,  
${{\tilde{T}}^r}{}_t$ is the electromagnetic part of ${T^r}_t$, 
$\epsilon = 1 - \langle\dot{E}\rangle/\langle\dot{M}\rangle \simeq 0.13$ is an 
effective radiative efficiency, 
$\langle\dot{M}\rangle \simeq 0.30$ is the average mass accretion 
rate through the horizon over $1000M<t<1500M$, and 
$\langle\dot{E}\rangle \simeq 0.26$ is the average 
energy accretion rate through the horizon over the same period. 
The mass flux efficiency is 
\beq{
\dot{m} = \frac{\dot{M}_\mathrm{jet}}{\langle\dot{M}\rangle} = 
\frac{2\pi}{\langle\dot{M}\rangle} \int_{d\theta_\mathrm{jet}} 
\rho_\circ u^r \sqrt{-g} d\theta
\quad .
\label{mass-flux-efficiency}
}
The electromagnetic luminosity component is significantly larger
for $r\lesssim100M$.  The increase in the luminosity fraction with radius is partially
due to collimation effects; the jet is wider than $15^\circ$ 
at smaller radii and collimates further out.
The similarity in the matter luminosity fraction and mass flux fraction is most likely 
from the jet's accumulation of mass from the floor.  We, however, still see the 
conversion of electromagnetic flux into matter energy flux seen by others 
\cite{2006MNRAS.368.1561M}.  
Let us assume that the free energy in our jet at large $r$ represents a 
reasonable estimate for the ultimate power of the jet.  We can then estimate 
the jet to have a luminosity of 
\beq{
L_\mathrm{jet} \approx 0.013 \dot{M} c^2  \quad . 
\label{jet-lum}
}
This value is similar to that calculated in other studies
\cite{2006MNRAS.368.1561M,2006ApJ...641..103H}, though each used
different floor schemes.

\section{Summary and Conclusion}
\label{sec:conclusion}

We have presented numerical estimates of the optically thin emission from 
GRMHD accretion disk simulations scaled to \sgr conditions  and commented
on the character of the jet seen in similar runs. 

Relativistic jets ($\gamma \lesssim10$) are seen from our geometrically 
thick accretion disks that remain collimated at large distances ($r\gtrsim1000M$).  
The energy flux is predominantly electromagnetic at small distances, but 
equipartition with the matter component is reached by $r\sim100M$. 
Our results are qualitatively similar to other studies 
\cite{2006MNRAS.368.1561M,2006ApJ...641..103H}. 

The ray-traced images 
of \sgr predict that the black hole silhouette will only be obvious 
near $\lambda\simeq1\mm$ if 
the disk is inclined less than $\sim30^\circ$ to the line of sight.
By taking pictures of the disks at different 
frequencies, we were able to calculate spectra for different inclinations, 
black hole spins and time slices.  Significant SED variations were seen with respect to 
all these parameters, though degeneracies may exist for certain combinations 
of parameters.  For instance, increasing $a_*$ and $i_\mathrm{inc}$ 
seemed to increase the power at high frequencies, so a low spin hole with a large
inclination angle may have a similar SED as a high spin hole with a small inclination
angle.  

Since the SED varies with time slice, we intend to take time averages of spectra 
to more accurately approximate real observations.  In addition, we plan on 
adapting our ray-tracing code to consistently calculate polarization through 
plasma on a curved background.  This will allow us to further constrain our model.  
Other improvements we 
plan to implement in the near future include removing the ``frozen fluid''
approximation which uses data from a single time slice to calculate the SED, 
adding Compton scattering and using 3D simulation data.

\ack{This research was supported by NSF grants AST 00-93091 and PHY 02-05155. }
\\



\end{document}